\documentclass[journal]{IEEEtran}
\usepackage{amsmath,amsfonts}
\usepackage{algorithmic}
\usepackage{algorithm}
\usepackage{array}
\usepackage[caption=false,textfont=sf]{subfig}
\usepackage{textcomp}
\usepackage{stfloats}
\usepackage{url}
\usepackage{verbatim}
\usepackage{graphicx}
\usepackage{cite}
\usepackage{color}
\usepackage{bm}
\usepackage[dvipsnames,table]{xcolor}
\usepackage{bbding}
\begin{document}
	
	\title{Collaborative Sensing in Perceptive Mobile Networks: Opportunities and Challenges}
	
	\author{Lei Xie,~\IEEEmembership{Member,~IEEE}, S.H. Song,~\IEEEmembership{Senior Member,~IEEE}, Yonina C. Eldar,~\IEEEmembership{Fellow,~IEEE}, \\
	and Khaled B. Letaief,~\IEEEmembership{Fellow,~IEEE}
		\thanks{L. Xie and S.H. Song are with Department of Electronic and Computer Engineering, the Hong Kong University of Science and Technology, Hong Kong. e-mail: ($\{$eelxie,eeshsong$\}$@ust.hk).}
		\thanks{Yonina C. Eldar is with the Faculty of Mathematics and Computer Science, Weizmann Institute of Science, Rehovot 7610001, Israel (e-mail: yonina.eldar@weizmann.ac.il).}
		\thanks{Khaled B. Letaief is with the Department of Electronic and Computer Engineering, the Hong Kong University of Science and Technology, Hong Kong, and also with Peng Cheng Laboratory, Shenzhen 518066, China (e-mail: eekhaled@ust.hk).}
	}
	\maketitle
	
	\begin{abstract}
		With the development of innovative applications that demand accurate environment information, e.g., autonomous driving, sensing becomes an important requirement for future wireless networks. To this end, integrated sensing and communication (ISAC) provides a promising platform to exploit the synergy between sensing and communication, where perceptive mobile networks (PMNs) were proposed to add accurate sensing capability to existing wireless networks. The well-developed cellular networks offer exciting opportunities for sensing, including large coverage, strong computation and communication power, and most importantly networked sensing, where the perspectives from multiple sensing nodes can be collaboratively utilized for sensing the same target. However, PMNs also face big challenges such as the inherent interference between sensing and communication, the complex sensing environment, and the tracking of high-speed targets by cellular networks. This paper provides a comprehensive review on the design of PMNs, covering the popular network architectures, sensing protocols, standing research problems, and available solutions. Several future research directions that are critical for the development of PMNs are also discussed. 
	\end{abstract}
	

	\section{Introduction}
	After several generations of development, wireless communications has evolved from a system with only communication services to an intelligent network that not only moves data but also performs edge-computing and distributed learning/inference tasks \cite{9606720}. The advancement of innovative applications such as autonomous driving and environment monitoring further requires accurate sensing capability from future wireless networks. To this end, the recently proposed integrated sensing and communication (ISAC) framework offers a promising way to integrate sensing and communication with possible hardware and software reuse, especially after millimeter wave (mmWave) was adopted for 5G and beyond systems. Perceptive mobile networks (PMNs) are a special type of ISAC system that focuses on adding sensing capability to the cellular networks \cite{9296833}. 
	
	There are many favorable properties of cellular networks that can facilitate sensing. First, the well-developed mobile network with high-density base stations (BSs) can provide large sensing coverage. The high-density nodes are very important because mmWave experiences severe pathloss and is thus not suitable for long-range sensing tasks. Second, the large number of distributed and connected sensing nodes enables networked sensing, where multiple perspectives from different sensing nodes can be exploited to sense the same target. Finally, the strong computation and communication power of PMNs create a good platform for large-scale environment estimation and mapping, which will not only benefit sensing but also enhance communication in terms of channel estimation, resource allocation, beam tracking, and more.    
	
	However, there are also challenges faced by the design of PMNs \cite{9296833}.  Since PMNs integrate sensing and communication in one system, interference management is one of the most important issues to tackle. In particular, there exist three types of interference. First, if the same node, e.g., a BS, is utilized for transmitting sensing/communication signals and receiving radar echoes at the same time, there will be self-interference (SI) \cite{9737357}. Second, given both communication and sensing users are served in the same frequency band, there will be interference between the two sub-systems. Finally, prior information about the environment is critical for sensing and normally obtained by environment training (estimation). In conventional radar systems, the transmitted signal in the environment training and target sensing periods is the same, thus guaranteeing the same covariance structure for the clutter. However, due to the high pathloss of mmWave, directional signals are utilized for probing a target in the sensing period, but not in the training stage. As a result, the sensing signal may cause different clutter covariance between the training and sensing periods, which can be regarded as the interference to environment estimation. 
	
	Besides interference management, the implementation of networked sensing and environment estimation algorithms also faces several obstacles. On the one hand, although networked sensing can take advantage of the multiple perspectives from several sensing nodes, the collaborative sensing algorithm must be computation and communication efficient due to the strict latency requirement. On the other hand, the environment for networked sensing is much more complex than traditional radar systems because the communication users, the target and even the clutter patches in the environment may move. This makes environment estimation very challenging, especially with distributed sensing nodes. Furthermore, the stringent latency constraint precludes estimation algorithms that require large amount of training data and heavy computation.                 
	
	This article aims to provide a comprehensive overview on the design of PMNs. For that purpose, we will first introduce and compare several existing network architectures and sensing protocols. In-depth discussions about the key research problems will then be given, revealing the key design opportunities and challenges in interference management, networked sensing and environment estimation. Future research directions that are critical for the development of PMNs and their service to other applications will also be covered. Different from the existing reviews about ISAC \cite{8999605,9127852,9345999,9737357}, this paper mainly focuses on PMNs especially the benefits of the networked sensing. 
	
	\section{Network Architecture and Sensing Protocol}
	Network architecture and sensing protocol are the two most fundamental frameworks for integrating sensing into current cellular networks.

	\begin{figure*}[!t]
		\centering
		\includegraphics[width=3.9in]{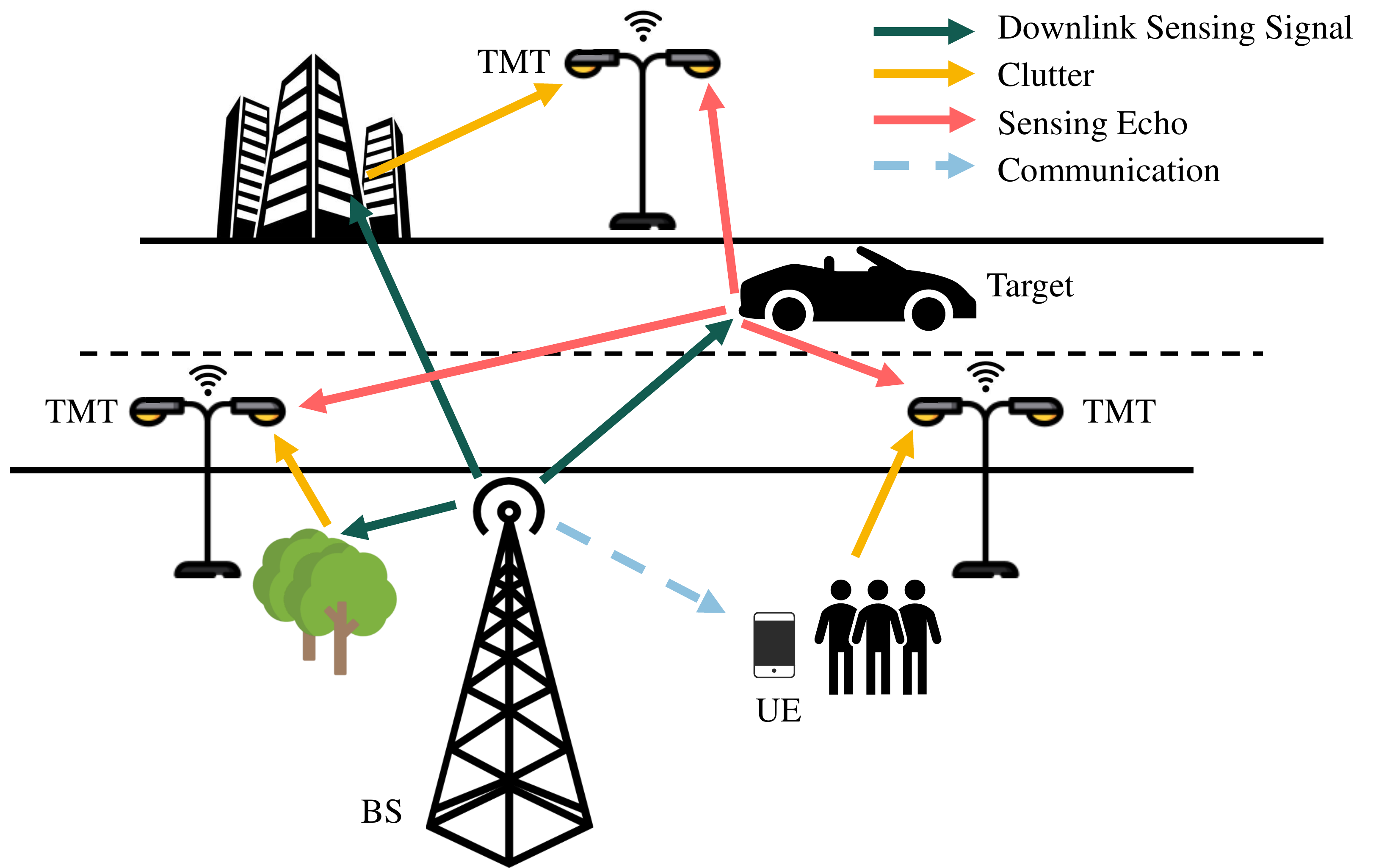}
		\caption{Illustration of the PMN-TMT architecture.}
		\label{sys1}
	\end{figure*}
	
	\begin{table*}[!htbp] 
		\centering 
		\caption{Comparison of existing ISAC networks.}
		\begin{tabular}{c|cm{3cm}<{\centering} m{3cm}<{\centering} m{3cm}<{\centering}}  
			\hline \hline
			\cellcolor[HTML]{D6DBDF}\quad 
			& \cellcolor[HTML]{D6DBDF} Sensing Transmitter 
			& \cellcolor[HTML]{D6DBDF} Sensing Receiver 
			& \cellcolor[HTML]{D6DBDF} Full-Duplex 
			& \cellcolor[HTML]{D6DBDF} Synchronization \\
			\hline
			\cellcolor[HTML]{EBEDEF} Mono-static DFRC 
			& \cellcolor[HTML]{EBEDEF} BS 
			&  \cellcolor[HTML]{EBEDEF} BS 
			& \cellcolor[HTML]{EBEDEF} Required 
			& \cellcolor[HTML]{EBEDEF} Not Required \\
			\cellcolor[HTML]{D6DBDF} PMN-RRU 
			& \cellcolor[HTML]{D6DBDF}RRU 
			& \cellcolor[HTML]{D6DBDF}RRU 
			& \cellcolor[HTML]{D6DBDF} Not Required  
			& \cellcolor[HTML]{D6DBDF} Required\\
			\cellcolor[HTML]{EBEDEF} PMN-TMT 
			& \cellcolor[HTML]{EBEDEF} BS 
			& \cellcolor[HTML]{EBEDEF} TMT 
			& \cellcolor[HTML]{EBEDEF} Not Required  
			& \cellcolor[HTML]{EBEDEF} Required\\
			\hline \hline
		\end{tabular}\label{Archtable}
	\end{table*}
	
	\subsection{Network Architecture Design}
	There are three main network architectures proposed for PMNs in the literature. In traditional cellular networks, interconnected BSs will serve mobile user terminals (UEs). It is thus natural to select the BSs as the transmitter for the sensing signal and the receiver for the echo signal \cite{8288677}. This is referred to as the mono-static dual-function radar and communication (DFRC) system, where the BSs are required to work in full-duplex (FD) mode for transmitting and receiving signals at the same time. Another architecture \cite{9296833} integrates sensing into the cloud radio access network (C-RAN) where remote radio units (RRUs) are densely distributed. To address the full-duplex issue, some RRUs are selected to be the dedicated receivers in the downlink sensing time, such that the sensing transmitter and receiver are separated to be different RRUs.  We will refer to this scheme as the PMN-RRU architecture. 

    A new PMN architecture was proposed in \cite{xie2021perceptive} where another layer of passive target monitoring terminals (TMTs) are added to the conventional cellular networks. TMTs are nodes designed for internet of things (IoT) applications with only passive sensing functionalities, such as radar and vision \cite{9143269}. They are distributed in a target area and connected with the BSs through low latency links. Given that TMTs will serve as dedicated radar receivers, BSs only need to serve as the transmitter for sensing signals, thus saving the need for full-duplex operation. We will refer to this design as the PMN-TMT architecture. 
	
	Both PMN-RRU and PMN-TMT avoid the full-duplex operation at the cost of network synchronization. Compared with RRUs, TMTs are low cost IoT devices with only passive sensing functions that can also be utilized for other types of IoT services. Fig. \ref{sys1} shows an illustrative example of the PMN-TMT scheme and Table \ref{Archtable} compares between the three architectures. Note that all three architectures may perform collaborative or networked sensing but with different sensing nodes, i.e., BSs, RRUs, and TMTs, respectively. In the following, we will simply use TMTs as the sensing nodes and all discussions apply to the other two architectures.
	
	\subsection{Protocol Design}
	
	ISAC systems have inherent interference issues and proper protocols play the key role in interference management and resource allocation between sensing and communication \cite{9127852}. In \cite{8999605}, the authors proposed a three-stage protocol to coordinate communication and sensing modules for a DFRC system. In Stage 1, the BS utilizes the omni-directional beam to search the sensing targets and scatterers, and receives the uplink pilot from the UEs. In Stage 2, the BS transmits the directional communication and sensing signals to the UEs and the sensing target, respectively, where the transmission directions are estimated in Stage 1. In Stage 3, the BS receives the radar echoes and the uplink signals from the UEs for target detection and tracking. A two-stage protocol was proposed in \cite{arxiv.2202.02688} to achieve joint target detection and channel estimation. In the initial stage, the BS sends omni-directional downlink pilots to search for the target, where the channel estimation is performed based on uplink pilots from the UEs. Based on the initial results about the target and communication scatterers obtained in the first stage, the BS sends directional downlink pilots in the second stage for refined target detection and channel estimation. 
	
	\begin{figure*}[!htbp]
		\centering
		\includegraphics[width=4.0in]{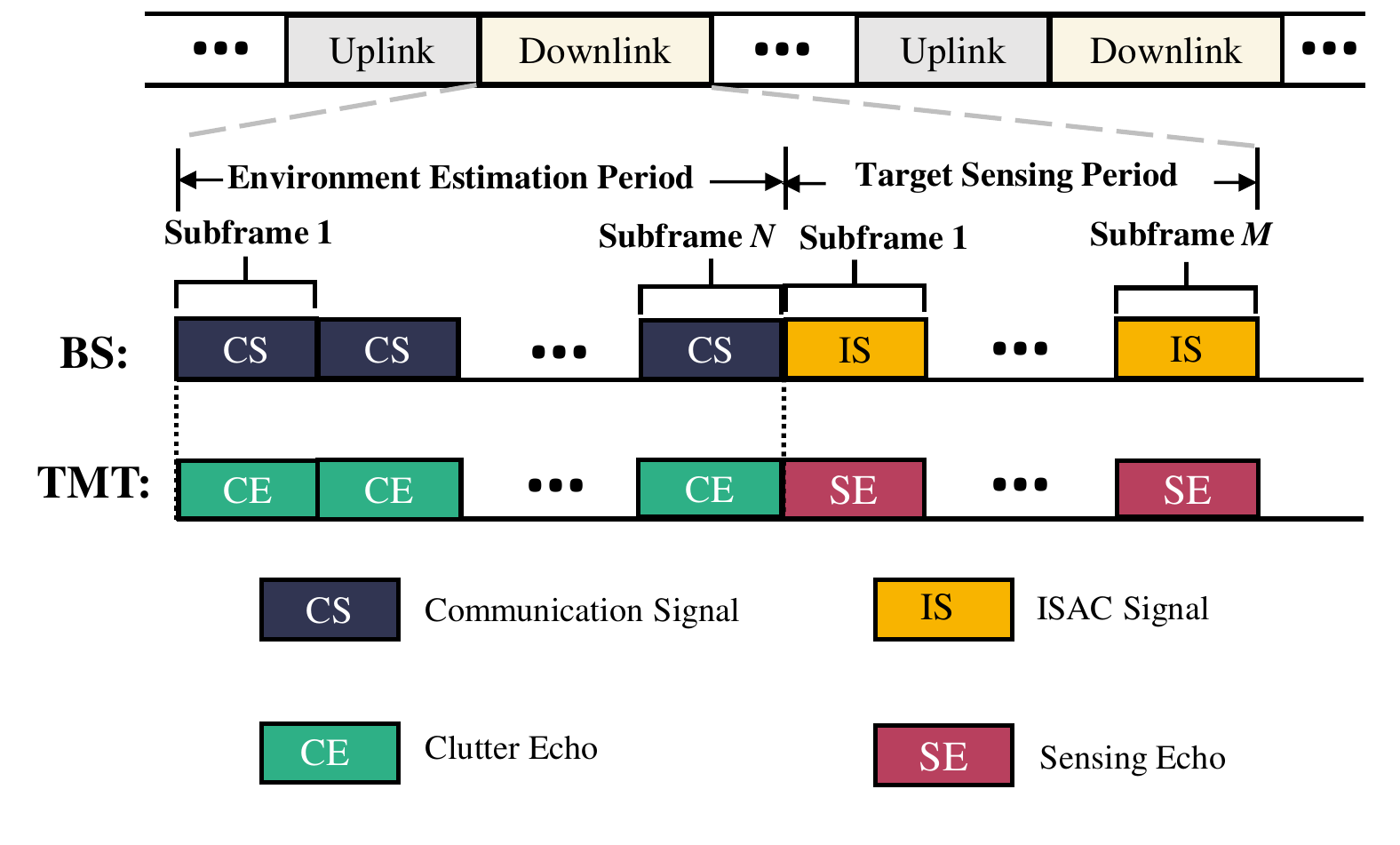}
		\caption{Frame structure for the two-stage sensing and communication protocol \cite{xie2022networked}.}
		\label{fig_framestructure}
	\end{figure*}
	
	To handle target sensing in complex environment with the presence of clutter, the authors of \cite{xie2022networked} proposed a two-stage sensing protocol where the communication signals are utilized to estimate the environment and sense the target in two consecutive periods as illustrated in Fig. \ref{fig_framestructure}. In particular, the downlink transmission time is divided into two periods, namely, the environment estimation (EE) period and the target sensing (TS) period. In the EE period, the BS transmits communication signals to serve the UEs, whose echoes from the environment will be captured by the TMTs to estimate the clutter covariance matrix. In the TS period, the BS transmits the designed ISAC signal to serve the UEs and probe a target simultaneously\footnote{The sensing signal is absent in the EE period and thus may change the environment statistics.}. TMTs receive the sensing echo (SE) reflected by the target and clutter patches, and perform target sensing, where the clutter is suppressed by exploiting the clutter covariance matrix estimated in the EE period.  Note that there are two interference management issues with this protocol: 1) the interference between sensing and communication; 2) the ISAC signal should not make the clutter covariance matrix in the TS period different from that in the EE period. See next section for more details. 
	
	\section{Key Research Problems: Opportunities and Challenges}
	The design of PMN is still in its infancy. In the following, we identify several key research problems and discuss the design challenges, opportunities, and existing solutions.
	
	\subsection{Interference Management}
	There are three types of interference in PMNs. 
	\begin{itemize}
		\item In DFRC, the BS needs to transmit and receive at the same time, causing SI. 
		\item As a special type of ISAC system, there is inherent interference between sensing and communication in PMNs.
		\item Due to the use of narrow beams in mmWave band, sensing signal toward a target may change the environment statistics, and this can be regarded as interference to environment estimation.
	\end{itemize}
	In the following, we discuss existing solutions for managing the above-mentioned interference.

	\subsubsection{Self Interference}
	Conventional pulse radar works in a half-duplex mode to avoid SI. In each pulse repetition period, antennas will utilize a long time to receive the potential echoes after transmitting the sensing signal. However, half-duplex mode  will not work for PMNs because BSs need to transmit communication signals all the time. Among the three architectures mentioned above, the DFRC scheme requires the BS to work in full-duplex mode to transmit and receive signals at the same time. Under such circumstances, the SI is mitigated by self-interference cancellation technique \cite{sabharwal2014band}, which unfortunately is not very mature. On the other hand, the PMN-RRU and PMN-TMT architectures naturally avoid SI.
	
	\subsubsection{Interference between Sensing and Communication}
	There are several unique features about sensing and communication signals in PMNs: a) the sensing signals will interfere with the UEs and may degrade the communication performance; b) the communication signals reflected by the target can be utilized to probe the target as they are known by the BSs \cite{8288677}; c) the communication signals reflected by the environment will cause clutter for sensing.  Depending on the design objective, different approaches have been proposed to manage the interference between sensing and communication. 
	
	There are works that consider both sensing and communication performance. Along this line, spectrum sharing between sensing and communication was investigated in \cite{li2016optimum} by dynamic spectrum access and mutual interference mitigation. 
	In \cite{xie2021perceptive}, the authors maximized the weighted average of sensing and communication performance by jointly designing the transmitter and receiver, where the sensing metric is selected as the signal-to-clutter-plus-noise ratio (SCNR) and the communication metric is the signal-to-interference-plus-noise ratio (SINR). An alternating optimization (AO) based framework is proposed to iteratively update the transmit and receive beamformers. There are also works that prioritize one side of the ISAC system. For example, the communication performance was maximized in \cite{zheng2017adaptive} by treating radar signals as interference.   
	
	In terms of the transceiver design, besides the AO-based method that jointly optimizes the transmitter and receiver, linear transceivers have also been considered to reduce the computation complexity. For example, zero-forcing (ZF) and beam synthesis (B-syn) transmitter, and the minimum variance distortionless response (MVDR) receiver were investigated in \cite{xie2021perceptive}. These linear transceivers not only reduce the computation complexity but also provide interesting physical insights. For example, ``leaking'' energy from communication signals to the sensing target (ST)\footnote{Here, leaking energy from communication signals means to revise the beamformer for the UEs so that part of the communication signals will also be sent toward the ST.} is more efficient than forming a dedicated sensing signal, and the amount of energy leaked from one UE to the ST depends on their channel correlation, which is determined by their locations.

	\begin{figure*}[!t]
		\centering
		\includegraphics[width=5.1in]{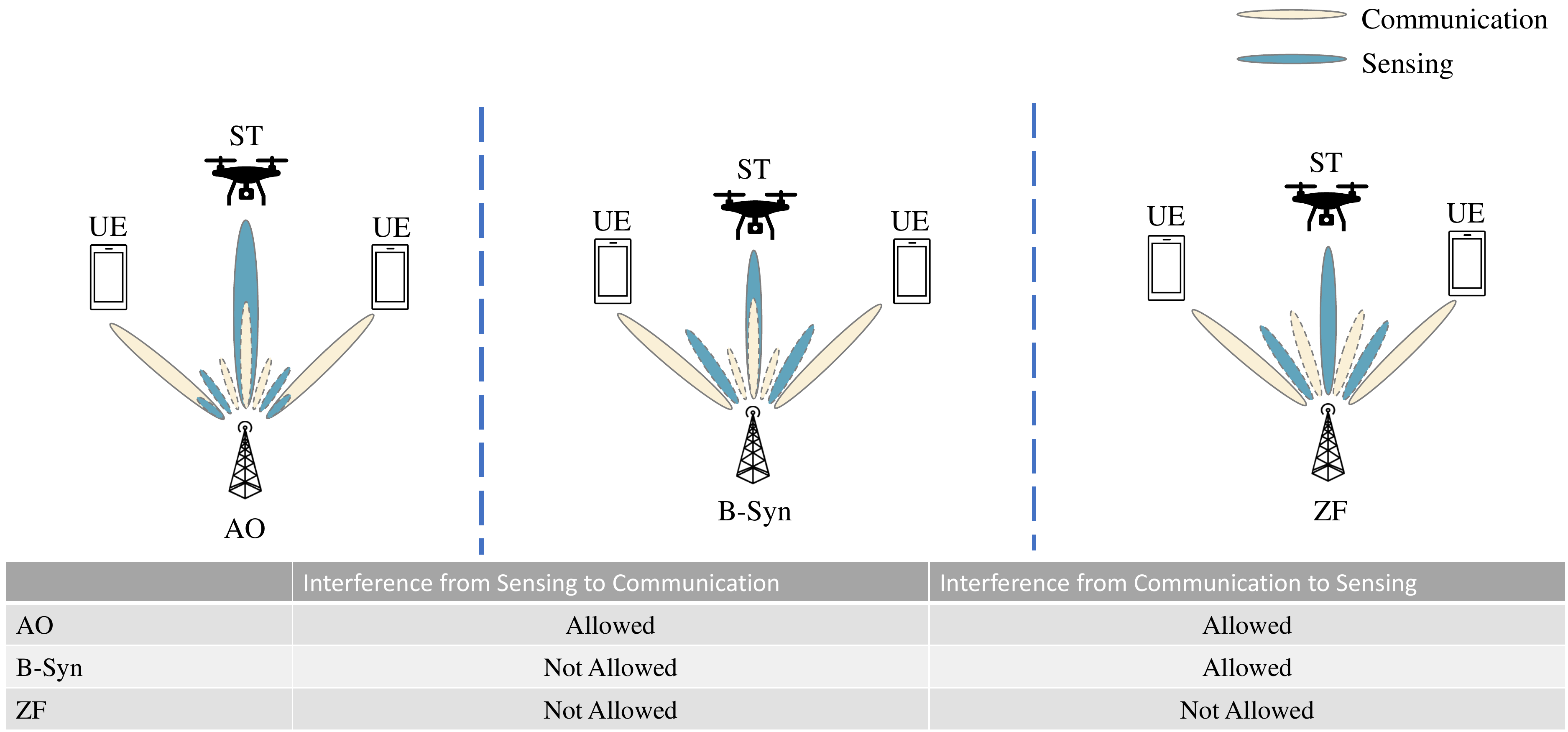}
		\caption{Comparison of the beam patterns for AO, B-syn and ZF.}
		\label{transceiver}
	\end{figure*}
	
	AO, B-syn and ZF represent three different ways to tackle interference between sensing and communication. In particular, AO allows interference between sensing and communication and handles it by joint transceiver design, B-syn only allows interference from communication to sensing and utilizes communication signals as the sensing signal, while ZF eliminates all interference between sensing and communication by designing a dedicated sensing signal that will not interfere with the UEs. The beam patterns of AO, B-syn, and ZF are illustrated in Fig. \ref{transceiver} where the interference between sensing and communication is also illustrated. 
	
\begin{figure}[!t]
		\centering
		\includegraphics[width=3.7in]{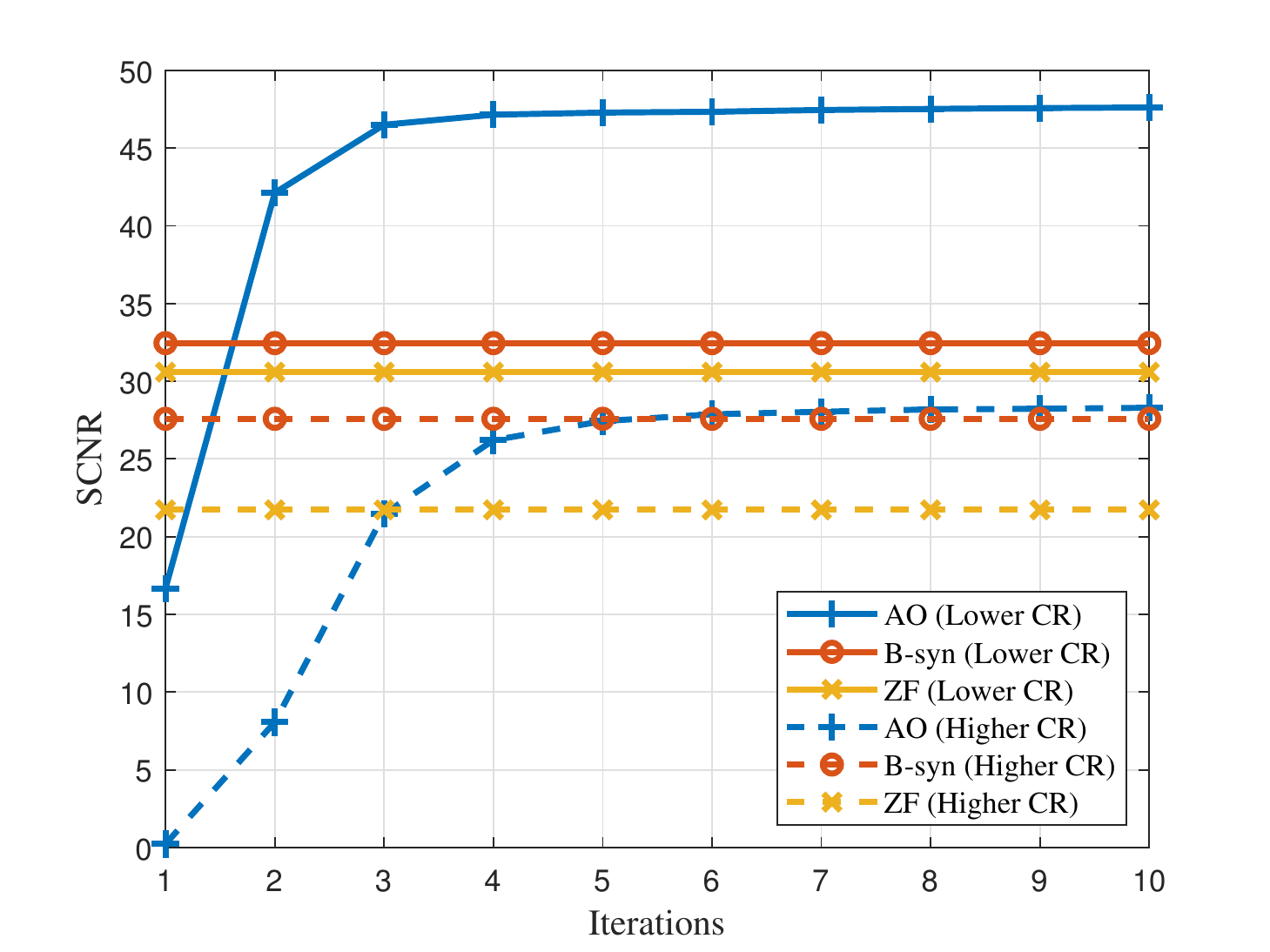}
		\caption{Detection performance comparison between AO, B-syn and ZF under different communication requirements (CR).
			The BS employs a ULA with $N_t = 16$ antennas. The receiver employs a hybrid beamforming structure with  $N_r = 16$ antennas and $N_{RF} = 4$ RF chains. The signal-to-noise ratio and clutter-to-noise ratio are $17$dB and $40$dB, respectively.}
		\label{fig_BT}
	\end{figure}

	Fig. \ref{fig_BT} compares the sensing performance of the three transmitters with high and low CR, respectively. It can be observed that, under both circumstances, the performance of AO is the best, and B-syn outperforms ZF. Note that different transceiver structures have different tolerance for interference, and the stronger the orthogonality constraint (ZF$>$B-syn$>$AO), the worse the sensing performance. Furthermore, when the CR is high, the gap between the three schemes is smaller. This is partially due to the less energy left for sensing when the CR is high. On the other hand, when the CR is low, high interference is acceptable to the UEs, thus it is not necessary to completely eliminate the interference from sensing to communication. However, high CR forces AO to avoid the interference from sensing to communication like what B-syn does. Therefore, the gap between AO and B-syn becomes smaller. When their performance is comparable, B-syn is preferred because its computational complexity is much lower than AO.

	\subsubsection{Interference to Environment Estimation}
	Besides the interference between sensing and communication, the sensing signal may also create interference for environment estimation. In particular, the transmitted signal in the TS period is different from that in the EE period, which may change the covariance structure of the clutter. To tackle this issue, the sensing signal in the TS period must be designed to avoid generating any echoes from the clutter patches, such that the covariance of the clutter received by TMTs will be the same as that in the EE period. For that purpose, the idea of B-syn can be adopted such that the sensing signal has no energy toward the clutter patches.  
	
	\subsection{Networked Sensing}
	Networked sensing is one of the most exciting opportunities for PMN, where the multiple perspectives from different TMTs can be collaboratively utilized to improve sensing performance. The biggest advantage of networked sensing comes from the macro-diversity of multiple TMTs. Note that, similar to wireless communications, the macro-diversity gain improves the reliability of target detection, because the chance that all the perspectives from different TMTs are blocked is very low. However, the impact of the multiple antennas at one TMT on sensing is totally different from that for wireless communication. The influence of both the number of TMTs and the number of antennas per TMT was analyzed in \cite{xie2022networked} with some interesting observations.  
	
	\subsubsection{Contribution of One Multi-antenna TMT}
	The contribution of a single multi-antenna TMT has several aspects. First, the distance between the ST and the TMT affects the receive SNR exponentially due to pathloss. Second, one TMT has its unique perspectives (AODs) toward the target and the clutter patches in the environment. The relation between those AODs dictates how easy it is for the TMT to separate the target from the clutter patches. Finally, the number of antennas at one TMT affects sensing in two aspects. On the one hand, the receive SNR at the TMT is directly proportional to the number of antennas, which is referred to as the antenna array gain. On the other hand, the number of antennas determines the array resolution, which indicates the TMT's ability to extract the target echo from the clutter. In particular, arrays with larger number of antennas can separate closer target and clutter patches. 
	
	However, multiple antennas no longer provide diversity gain for sensing. Note that multi-antenna receivers achieve diversity gain in wireless communications because the channel coefficients of different transmitter-receiver antenna pairs are not fully correlated or even independent. However, in sensing, only the line-of-sight (LoS) component is considered, and all non-LoS components are regarded as part of the clutter. As a result, the channel between the ST and the TMT becomes deterministic and no diversity can be achieved by the multiple antennas of one TMT. 
	
	\subsubsection{Macro-diversity and TMT selection}
	Due to the different perspectives and independent reflecting coefficients, the channels between the ST and different TMTs are normally independent. As a result, multiple TMTs will provide macro-diversity gain for sensing. But, this doesnot mean that we should include as many TMTs as possible, because the detection probability is not a monotonic increasing function of the number of TMTs.  This phenomenon was analyzed and discussed in \cite{xie2022networked}. Assume there are already $L$ TMTs participating in the networked sensing. A new TMT, i.e., the ($L$+1)-th TMT, will change the distributions of the decision statistic under both the target-absence (TA) and target-presence (TP) hypotheses. Consider an extreme case where the new target-TMT link is blocked. Under such circumstances, the ($L$+1)-th TMT will only contribute noise under both hypotheses, and thus cause worse performance. As a result, it is unnecessary and even harmful to activate all TMTs to sense one target, making TMT selection a critical task. In \cite{xie2022networked}, a sufficient condition for the contribution of one more TMT to be positive was derived, with which a TMT selection algorithm was proposed. 
	
	\subsection{Environment Estimation}
	
	Another challenge for PMNs is EE. To achieve good sensing performance in the presence of clutter, PMNs need to suppress the clutter by utilizing the prior information about the environment obtained by EE. The performance of clutter suppression will degrade if the clutter component of the signal-under-test during the TS period has different statistical structure from that of the EE period.  A compressed sensing (CS)-based method was proposed in \cite{8827589} to estimate the spatial parameters and Doppler shift of the clutter. Unfortunately, the computational cost of the CS-based method can be extremely high, due to the continuous and rapidly-changing environment parameters in the space and Doppler domains. This issue will be more serious when multiple distributed TMTs are involved in the networked sensing where information sharing between different TMTs is necessary. Thus, a computation and communication efficient EE algorithm is desired. 
	
	To reduce the computation and communication workload for EE, the authors of \cite{xie2022networked} proposed a distributed clutter covariance estimation algorithm where the estimation is performed at TMTs. The low rank clutter in the mmWave band \cite{li2016optimum} makes it possible to estimate the clutter covariance by using partial samples of the received signal. However, the estimated covariance matrix may be ill-conditioned due to the limited data samples. To this end, the EM-Net algorithm was proposed by unfolding the expectation-maximization (EM) detector with several learnable parameters, which achieves accurate estimation with less data samples than existing methods.
	
	\begin{figure}[!t]
		\centering 
		\includegraphics[width=3.7in]{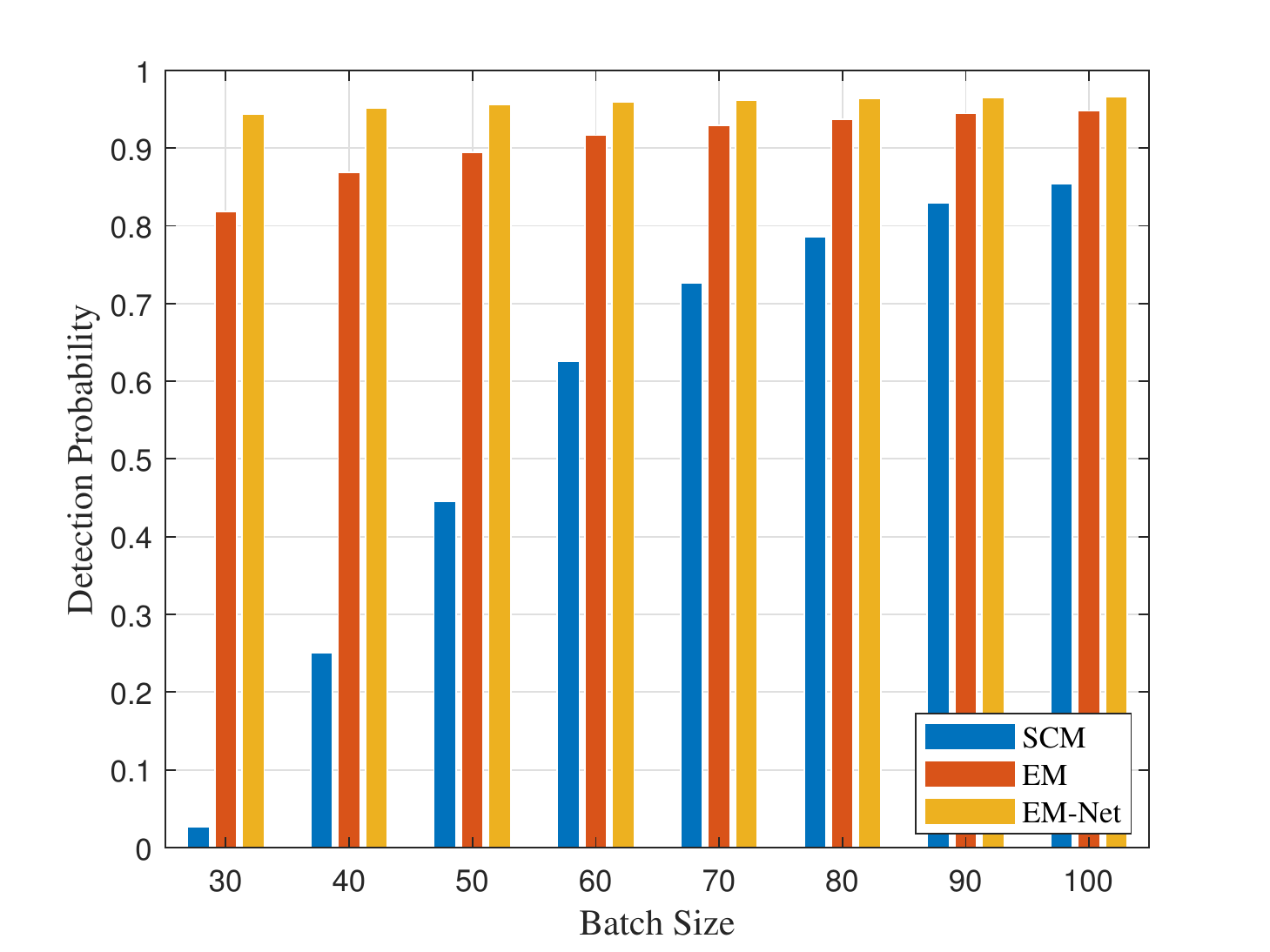}
		\caption{Detection performance versus the batch size for different environment estimation methods. The number of the antennas at one TMT and the number of TMTs are set as $N_R = 16$, and $L=3$, respectively.}
		\label{fig_batch}
	\end{figure}

	Fig. \ref{fig_batch} shows that EM-Net outperforms both the EM approach and the conventional sample covariance matrix (SCM) based method, and the detection performance will improve as the batch size for EE increases.  However, the performance improvement is achieved at the cost of larger latency and higher computation workload.
	
	\section{Future Research Directions}
	In this section, we list several future research directions for the development of PMNs. 
	
	\subsection{Target Tracking}
	A big challenge for networked sensing is to track a moving target. This is because, different from traditional radar, the coverage of one TMT is very limited. Thus, the sensing task needs to be smoothly handed over to another TMT when the target is moving. The situation is even worse for networked sensing where the handover is not from one TMT to another TMT, but from one group of TMTs to another group with overlapping between two groups. Note that handover is not a new problem, as it is also required for wireless communication services. However, the problem is not the same for sensing due to several reasons. First, the synchronization requirement for networked sensing makes the problem much more difficult. Second, the handover criteria are more complex. Specifically, besides the signal strength which is normally the criterion for handover in communication, angles (AOA/AOD) and moving direction also matter a lot for sensing. Finally, if the target is moving fast, the narrow sensing beams may have difficulty following the target so that ``predictive'' sensing beams may be required.  
	
	To this end, besides the Kalman filter (KF) based method, machine learning or data-driven approaches may play a more important role due to several reasons. On the one hand, the physical environment, e.g., the roads, for fast moving target is relatively fixed, making it easier to extract useful information from history data. On the other hand, the strong computation and communication power of PMNs make it possible to build a map of the environment, i.e., Simultaneous Localization and Mapping (SLAM). Furthermore, graph neural networks (GNNs) have been shown effective to handle wireless network design problems, such as traffic prediction, resource allocation and data detection, and the handover problem can also be formulated as a graph optimization problem.
	
	\subsection{Joint Networked and Individual Sensing}
	Networked sensing can be utilized to facilitate the design of many smart applications. For example, in autonomous driving, a key task for the vehicle is to understand the environment, including the road, other vehicles, pedestrians, etc. Currently, this is mainly achieved by many sensors installed on the vehicle, such as vision, lidar, and radar. We will refer to this as individual sensing. Networked sensing by PMNs can help measure the position and velocity of a moving vehicle. Due to latency requirement, such results can not be directly utilized to control the vehicle. However, networked sensing may provide useful complement for individual sensing in autonomous driving applications.
	
	There are several promising research directions. First, networked sensing is able to reduce the workload of individual sensing. For example, networked sensing can build up and keep updating a map about the static environment, saving the need for individual sensing to construct such a map. Second, networked sensing will be able to monitor the changing environment, e.g., other moving vehicles, to provide assistant information for individual sensing. Finally, many long-term tasks such as routing and traffic management can be taken over by networked sensing.

	\subsection{Intelligent Reflecting Surfaces (IRS)-aided Sensing}
	Compared with the sub-6GHz band, mmWave experiences high pathloss which makes directional transmission by spatial beamforming inevitable. As a result, both mmWave sensing and communication rely heavily on the LoS link, which unfortunately can be easily blocked. Intelligent reflecting surfaces (IRSs) have been proposed to create an alternative link between the transmitter and receiver, and attracted much attention in wireless communication design. The ability IRSs in creating semi-LoS links can be utilized to facilitate several aspects of PMN design including interference management, networked sensing, and velocity estimation. In particular, due to the complex environment, it is possible that the sensing target has the same AoA/AoD as some UEs with respect to one TMT. Under such circumstance, IRSs will be able to create another link and avoid the interference between sensing and communication users. The same idea can be utilized for networked sensing, when one TMT's perspective to the target is blocked. In fact, with proper phase-shift design, one IRS can help multiple TMTs. 
	
	The application of IRS in sensing is not limited to creating alternative paths. The additional perspective can also help improve velocity estimation. For example, with the conventional mono-static radar, only the radial projection of the true velocity can be estimated due to the nature of Doppler effect. As a result, if the target is moving on the direction perpendicular to the line connecting the target and one TMT, the velocity estimation will not be accurate. The additional path provided by the IRS can provide another perspective to observe the target, which is useful to recover the true velocity.

	\subsection{Joint Active and Passive Sensing}
	The above discussions considered networked sensing in an active manner, i.e., the BS actively transmits the sensing signal and TMTs perform the detection based on the echoes. The active mode works for any targets. However, if the PMN wants to locate the UEs which also use the communication service, there is another mode of passive sensing.  In particular, the UEs will transmit signals in the uplink period, which can also be used for radar detection. For targets with weak electromagnetic wave reflection characteristics, e.g., pedestrians, the passive method can provide a good detection performance. The passive mode is also more power-efficient as there is no transmission overhead. However, it loses several advantages of active sensing, such as the well-designed waveform, flexible transmitting beam, higher detection range, etc. As a result, it may not be sufficient to only utilize the passive mode to achieve accurate sensing performance. A joint design between active and passive sensing in PMNs is an interesting direction to investigate, which is especially useful for sensing-aided communication. 
	
	\subsection{Sensing-aided Communication}
	By far, we have been focusing on the design of PMNs to achieve sensing by exploiting the well-developed communication network. On the other hand, sensing can also be utilized to assist communication especially in the mmWave band where highly directional signals are transmitted in a sparse channel.

	\subsubsection{Sensing-aided Channel Estimation}
	The transmission of highly directional beams over a sparse channel creates challenging problems for channel estimation, because of the very large searching space in the angular domain plus the limited measurements. Among other solutions, the CS-based methods were widely studied to exploit the sparse structure, but they suffer from high computational complexity and are not robust to noise, hardware-led errors in array response, and the off-grid issues. To this end, the sensing results by PMNs can be of great assistance. In particular, the sparse channel is composed of several main scatterers whose locations can be obtained by sensing. As a result, sensing results can significantly reduce the searching space for channel estimation and improve the performance with limited measurements. 
	
	\subsubsection{Sensing-aided Beam Tracking}
	Beam alignment is a fundamental issue in mmWave communication. The narrow mmWave beams are very sensitive to the change of environment, e.g., the movement of the UEs. As a result, beam alignment becomes more difficult in highly mobile scenarios, as the state of a UE can change before beam training has been completed. Thus, under certain circumstance, ``predictive'' beams are required to maintain good communication performance. To this end, the powerful capability of sensing in location and velocity estimation will be able to help for tracking the moving UE and forming communication beams. 
	
	\section{Conclusion}
	Networked sensing brings unprecedented opportunities to exploit the well-developed infrastructure of cellular networks for sensing purposes, but at the same time faces serious challenges in interference management and environment estimation. Joint processing among distributed nodes over the network also incurs difficulties in designing communication and computation-efficient algorithms. Existing network architectures, sensing protocols, and transceiver design could tackle some of the challenges while achieving favorable results such as the macro-diversity from multiple sensing nodes, the array gain by multiple receive antennas, and the efficient environment estimation with data-driven methods.  However, the development of ISAC/PMNs is in its infancy and there are still many obstacles to conquer before we can fully enjoy the synergy between sensing and communication to support more innovative applications.

\end{document}